\documentclass[11pt]{article}
\usepackage[utf8]{inputenc}
\usepackage{lipsum}
\usepackage[square,sort,comma,numbers]{natbib}
\usepackage{natbib}
\usepackage{hyperref}
\usepackage{enumitem}
\usepackage{float}
\usepackage{graphicx}
\usepackage{caption} 
\usepackage{calc}
\usepackage{fancyhdr}
\usepackage{tcolorbox}
\setcitestyle{square}
\usepackage{hyperref}
\usepackage{wrapfig}
\usepackage{graphicx} 
\usepackage{amssymb}
\usepackage{amsmath}
\usepackage{multicol}
\usepackage{braket}
\usepackage{abstract}
\usepackage{multicol}
\title{\vskip-.7cm \large \bf Mass Superinflation from Enhanced Cauchy Horizon Singularity 
\\in a Reissner-Nordstr\"om Black Hole}

\author{\normalsize \bf  Nihar Ranjan Ghosh\footnote{g.nihar@iitg.ac.in}\ \ and Malay K. Nandy\footnote{mknandy@iitg.ac.in}\\
Department of Physics, Indian Institute of Technology Guwahati\\
Guwahati 781 039, India}

\date{January 20, 2024}

\begin{document}
\maketitle

\begin{abstract}
Ever since Penrose and Simpson contradicted Novikov’s prediction that an infalling passenger would emerge into an asymptotically flat universe, there have been a continued interest in predicting the nature of singularity at the Cauchy horizon of a Reissner-Nordstr\"om  black hole. This prediction was first confirmed by Poisson and Israel using cross-stream of massless particles, suggesting the phenomenon of \textit{mass inflation}. Ori however obtained a weaker singularity using a null shell of radiation. 
In this work, we consider a massive scalar field coupled to the Reissner-Nordstr\"om geometry and analyze the nature of singularity at the Cauchy horizon. To study the asymptotic behavior of the mass function and the scalar field near the
Cauchy horizon, we perturbatively solve the coupled dynamical equations emplyoing the Adomian decomposition method.
Our analysis shows that the mass function exhibits a very rapid and unbounded double-exponential growth, called herein \textit{mass superinflation}, which is enormously stronger than previously obtained singularities. The scalar field is also found to undergo a very strong blueshift  near the Cauchy horizon. 

\end{abstract}

\begin{multicols}{2}

\section{\label{sec:level1}Introduction}

The result of any general gravitational collapse is a trapped region, which indicates the presence of both an outer and an
inner apparent horizons. Classically the outer apparent horizon shifts or moves outward (as more mass falls into the black hole), while the inner one moves inwards, either settling to
a final constant radius or collapsing all the
way to the origin and creating a spacelike singularity. If the inner horizon tends to a final position of constant radius then it leads to the well-known mass
inflation instability phenomenon, where any perturbation in the
matter content of the geometry leads to the increase of curvature at the Cauchy horizon (CH). The most common geometries with an inner horizon are the Kerr and Reissner-Nordstr\"om solutions. In this work we study the behaviour at the CH in the Reissner-Nordstr\"om (RN) black hole.

 A Reissner-Nordstr\"om spacetime is formed when an electrically charged spherically symmetric mass undergoes sufficient gravitational collapse. The static geometry of the spacetime is obtained as a unique solution of the Einstein-Maxwell equation \cite{reissner1916eigengravitation},\cite{nordstrom1918energy} and the spacetime metric is described by 
 \begin{equation}
\label{metric in t,r coordinate}
    ds^2=-f(r) dt^2+ \frac{dr^2}{f(r)}+r^2 d\Omega^2~.
\end{equation}
where $f(r)=(1-\frac{2m_0}{r}+\frac{Q^2}{r^2})$ is the redshift function, with $m_0$ the ADM mass and $Q$ the electric charge of the black hole (BH) and $d\Omega^2=d\theta^2+\sin^2\theta d\phi^2$. In this geometry, there are two horizons, given by $f(r)=0$,$$
r_{\pm}=m_0\pm\sqrt{m_0^2-Q^2}~.
$$ The inner horizon $(r_-)$ of the Reissner-Nordstr\"om black hole also serves as the Cauchy Horizon, because the zero of the redshift function has a negative slope at the inner horizon. Hence the inner horizon acts as a boundary for predictability. The cosmic censorship conjecture implies $m_0\ge Q$, with the equality (inequality) giving extremal (non-extremal) black hole \cite{penrose1969gravitational}. 

\begin{figure}[H]
  \centering
    \includegraphics[scale=0.3]{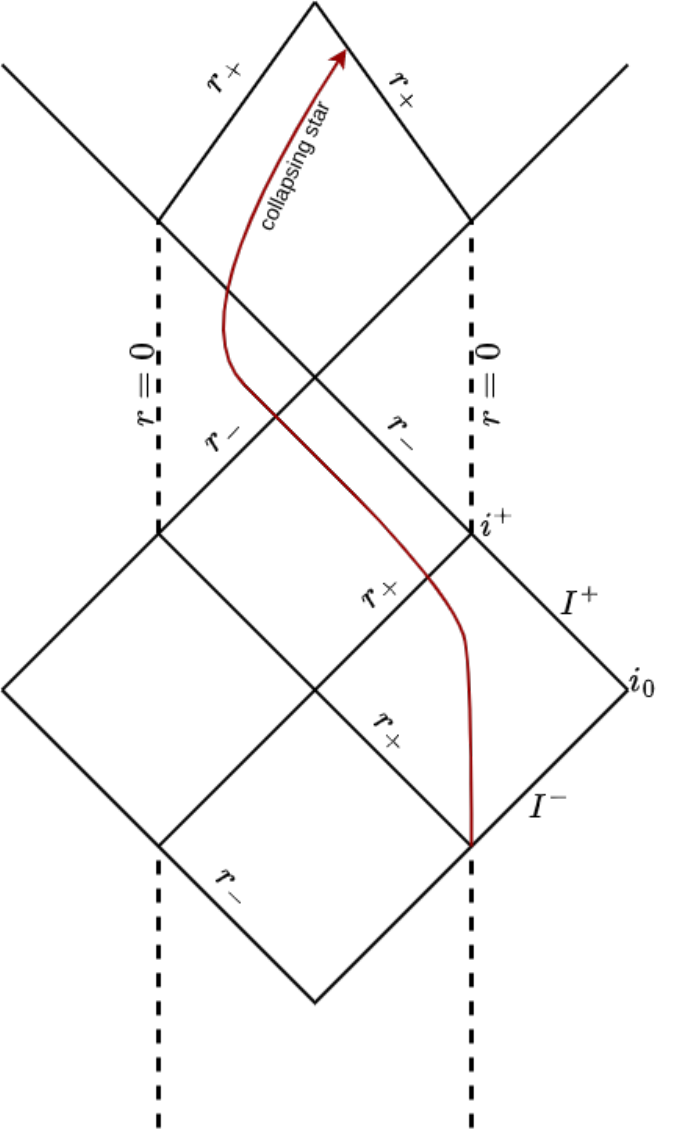}
    \caption{Penrose diagram of a Reissner-Nordstr\"om black hole showing the trajectory of an infalling passenger who manages to skip the central singularity and emerges into another universe.}   
    \label{figp}
\end{figure}

The extended Penrose diagram for non-extremal Reissner-Nordstr\"om  black hole with an infalling passenger in this background is shown in Figure \ref{figp}. 
The redshift function, $f(r)$, changes sign in the region $r_-<r<r_+$, which implies that the roles of $t$ and $r$ coordinates are interchanged.

The presence of a timelike singularity gives rise to a strange phenomenon. Novikov \cite{novikov1966change} first argued that an object falling into an RN black hole can emerge into an asymptotically flat universe without hitting the central singularity. During gravitational collapse with a given initial condition, the internal dust evolves in a completely regular manner, creating a \textit{tunnel} to another universe, as shown in Figure \ref{figp}.

However, this deterministic behavior is avoided by the instability of the inner horizon, serving as a barrier to predictability. Simpson and Penrose \cite{simpson1973internal,Schpf1970BattelleR}, shortly after the formulation of the singularity theorem, considered an infalling test electromagnetic field in the RN spacetime. They confirmed that the (CH) becomes unstable as the test field develops a singularity. Thus an infalling passenger will not survive upon encountering the CH. 

Penrose's prediction of CH instability can be understood by considering an infalling test field \cite{ori1991inner} such as a wave with uniformly spaced crests, which will become densely packed and experience gravitational and Doppler blueshifts. This amplification continues indefinitely at the CH signifying its instability to any external perturbation \cite{hamilton2010physics}. Owing to the backreaction of this divergence, the regular CH transforms into a curvature singularity.
After Penrose, several confirmations were made regarding this divergence up to linear perturbation, but no analysis was carried out beyond linear order until Poisson and Israel.
\subsection{ Classical Mass Inflation}
Poisson and Israel \cite{poisson1989inner}, first examined the behavior of the mass function accounting for the back reaction on the CH in the RN geometry due to infinitely blueshifted infalling spherically symmetric stream of massless particles. This resulted in a charged Vaidya geometry and, for a suitable choice of coordinates, the singularity was weak.
However, upon including the backscattered outgoing stream, the 
mass function behaved as $
m(U,V)\sim m_0\epsilon^2\delta(U)e^{k_0v}(v/m_0)^{-12}
$. Thus within a time scale of $\Delta v\sim k_0^{-1}\sim m_0$, the curvature $m/r^3$ must reach Planckian values near the CH (at constant $U$), where $v$ is the retarted time, $\epsilon$ is the quadrupole moment, $\delta(U)$ is the fraction of mass radiated outward between the moment of entry and time $U$. The outgoing radiation is the main reason behind this sudden growth in strength of the singularity, as was first found by Poisson \cite{poisson1990internal}.

In this process the mass function diverges exponentially (coined as \textit{mass inflation}) and the Kretschmann scalar
 $R_{\mu\nu\alpha\beta}R^{\mu\nu\alpha\beta}$ approaches infinity at the CH. This is a stronger singularity than the charged Vaidya solution. The reason behind this strong singularity is the failure of the cross streams to move faster than light with respect to each other, which produces a large amount of gravitational energy that reveals itself as an infinite amount of local mass near the CH \cite{brown2011mass}. 
 
Ori \cite{ori1991inner} improved the analysis by patching two Vaidya solutions separated by an infalling null shell of radiation, yielding
$m(v)\approx\Delta m(v)\propto|v|^{-1}(-\ln|v|)^{-p} $, where $p\ge12$ from Price's law\cite{price1972nonspherical}. Thus $m(v)$ and $R_{\mu\nu\alpha\beta}R^{\mu\nu\alpha\beta}\propto m^2$ diverge near the CH.

 Ori showed that the growth rate of curvature and tidal forces is proportional to $\tau^{-2}|\ln|\tau||^{-p}$, $\tau$ being the proper time and particle separations constituting an infalling passenger remain finite as $\tau\rightarrow0$, so that the spacetime can be extended beyond the CH despite the (weak) mass inflation singularity. Singularity strength criteria by Tipler \cite{tipler1977singularities}, Ellis and Schmidt \cite{ellis1977singular}, and others \cite{burko1998analytic, dafermos2017interior} confirm the same conclusion.

 Consequently attempts have been made in semiclassical frameworks to address this problem.
\subsection{Semiclassical Mass Inflation}
Employing the semiclassical treatment of Davies and Fulling \cite{davies1976energy} for the stress energy tensor, Hiscock \cite{hiscock1977stress} first showed that the stress energy tensor diverges at the Cauchy horizon. 
Moreover, Balbinot and Poisson \cite{balbinot1993mass} considered the semiclassical effect \cite{birrell1984quantum} on the mass inflation including quantum corrections to the stress energy tensor. Although they did not give a definite conclusion, they speculated that the mass inflation singularity may become stronger or regular depending on the sign of the quantum correction. 

Hwang and Yeom \cite{hwang2011internal} examined the interior structure during the formation and evaporation of RN black holes in the semiclassical formulation. Employing S-wave approximation they obtained a wormhole-like internal structure when $Q/M$ is small. However, in the final stages,  mass inflation generates a large curvature in the interior.

Recently Barcel\'o \cite{barcelo2021black, barcelo2022classical} analyzed the backreaction of a quantum field on the RN geometry. They incorporated, in the Polykov approximation, the renormalized stress energy tensor of a massless quantized scalar field and found that the classical tendency of the inner horizon to move inward may be overcome by the semiclassical tendency to move outward. 

Carballo-Rubio et al. \cite{carballo2022regular} and Franzin \cite{franzin2022stable} took regularized black holes and argued that they have vanishing inner horizon surface gravity. However, Maken \cite{mcmaken2023semiclassical} proved that the CH instability still continues to exist.
\subsection{de Sitter Models}
Motivated by the accelerated expansion of the Universe, Brady and Poisson \cite{brady1992cauchy} modeled the RN BH in de Sitter space, perturbed by an infalling stream of particles. Consequently the CH instability occurs when its surface gravity exceeds that of the cosmological horizon. However, for a generic perturbation, the CH remains stable, requiring $|Q|>M$ \cite{brady1992cauchy, mellor1990stability}.

Brady et al. \cite{brady1993cauchy} showed for a perturbed RN BH that for specific region in the parameter space the mass function remains bounded while the curvature scalar diverges.

Cai and Su \cite{cai1995black} employed the stability conjecture of Helliwell and Konkowski \cite{helliwell1993testing, konkowski1994instabilities}, confirming the prediction of Brady et al. However, Markovi\'c and Poisson \cite{markovic1995classical} claimed that the CH must be unstable when treated quantum mechanically.

\subsection{Present Work} 
We see from the above discussions that the problem of mass inflation has been attacked from various angles in both classical and semiclassical frameworks. All these studies lead to one common feature that the singularity of the CH continues to exist both classically and semiclassically. One important feature which remains still unclear is that whether the CH instability is strong enough  so that the possibility of extending the spacetime beyond the CH does not exist. Another common feature that can be noticed is that those calculations were based on ingoing and outgoing null shells of massless radiation. A more physical approach would be to consider a simple but realistic model describing the scenario by a \textit{massive scalar field} as astrophysical black holes are normally surrounded by massive particles in addition to massless radiation.

In this paper, we therefore consider a  massive, chargeless classical scalar field in spherically symmetric Reissner-Nordstr\"om geometry and study the behaviour of the singularity at the Cauchy horizon. We find that the singularity is enhanced double exponentially as the CH is approached asymptotically, where the scalar field undergoes an extraordinary blueshift.  These asymptotic behaviours are much stronger than those in previous studies and they suggest that the  inner horizon serves as an impenetrable physical boundary of the spacetime.

The rest of the paper is organised as follows. In Section \ref{sec-2}, we introduce our model of massive scalar field in the Reissner-Nordstr\"om geometry and obtain the Einstein-Maxwell field equations coupled with dynamics of the massive scalar field. In Section \ref{sec-3}, we develop a perturbation theory to solve the coupled set of differential equations which gives the solutions in terms of an integro-differential equation. In Section \ref{sec-4}, we give a brief account of the Adomian decomposition method (ADM) to be utilised in Section \ref{sec-5}, where we employ this method to solve the equation obtained in Section \ref{sec-3} in the asymptotic limit. Finally in Section \ref{sec-6}, we give a discussion and conclusion of our results.

\section{The Model}
\label{sec-2}
As discussed above, we study the behaviour of the CH singularity by perturbing the black hole with a massive, chargeless, scalar field $\Phi$. Thereby, we construct the  dynamics of the scalar field coupled with the dynamics of the mass function. In order to describe mass inflation, we shall solve the coupled dynamical equations in the near horizon and long time limits. 

We begin our calculation by modeling the phenomenon with the combined action 
\begin{equation}
    \label{action}
   S=S_g+S_\Phi~,
\end{equation}
with 
\begin{equation*}
    S_g=\int d^4x \sqrt{-g}\left[\frac{1}{2}M_P^2 R-\frac{1}{4}F_{\mu\nu}F^{\mu\nu}    \right]~,
\end{equation*}
and 
\begin{equation*}
     S_\Phi
     =\int d^4x \sqrt{-g}\left[-\frac{1}{2}g^{\mu\nu}\nabla_\mu\Phi\nabla_\nu\Phi-\frac{1}{2}\mu^2\Phi^2\right]
\end{equation*}
where $R$ is the Ricci scalar, $\mu$ is the mass of the scalar field $\Phi$ and $F_{\mu\nu}$ is the electromagnetic tensor.  The spacetime metric $g_{\mu\nu}$ is spherically symmetric in the Reissner-Nordst\"om geometry and it is given by the line element  
\begin{equation}
    \label{metric in tortoise coordinate}
    ds^2=f[-dt^2+dr^{*2}]+r^2 d\Omega^2~,
\end{equation}
where the tortoise coordinate $r^*$ is defined by $dr^*=dr/f(r)$. For further convenience, we rewrite the line element in $(v,r)$ coordinateS as
\begin{equation}
\label{metric in v,r coordinate}
    ds^2=-f dv^2+2dv dr+r^2 d\Omega^2~.
\end{equation}

Now, because of the perturbation by the massive scalar field, the mass  $m_0$ will be modified to $m_0+m(v)$, so that $$f=1-\frac{2[m_0+m(v)]}{r}+\frac{Q^2}{r^2}~.$$
Henceforth, we shall denote the inner horizon ($r_-$) by $r_0$. 

Equipped with the above formulation, we obtain from the action (\ref{action}), the Einstein field equation, 
\begin{eqnarray}
    \begin{split}
         \label{Einstein}
     R_{\mu\nu}-&\frac{1}{2}g_{\mu\nu} R =\frac{1}{M_p^2}\left[F_{\nu\beta}F_\mu^\beta- \frac{1}{4}g_{\mu\nu}F_{\alpha\beta}F^{\alpha\beta} \right.\\
    & \left.+\nabla_\mu\Phi\nabla_\nu\Phi-g_{\mu\nu}\left(\frac{1}{2}\nabla^\rho\Phi\nabla_\rho\Phi+\frac{1}{2}\mu^2\Phi \right)    \right]~. 
    \end{split}
\end{eqnarray}
The $00$ component of the above equation yields the dynamics of the mass function as 
\begin{equation}
    \begin{split}
        \label{mass}
    \frac{d m}{d v} =4\pi r^2& \left[   \left(\frac{\partial\Phi}{\partial v}  \right)^2+f\left(\frac{\partial\Phi}{\partial v}  \right)\left(\frac{\partial\Phi}{\partial r}  \right)\right.\\
    &+\left.\frac{1}{2}f^2\left(\frac{\partial\Phi}{\partial r}  \right)^2+\frac{1}{2}f\mu^2\Phi^2 \right]~.
    \end{split}
\end{equation}

Moreover, the action (\ref{action}) gives the Klein-Gordon equation for the massive scalar field, $\left[\square-\mu^2\right]\Phi=0$, which can be written as
\begin{equation}
    \label{KG}
    \begin{split}
        2r^2\frac{\partial^2\Phi}{\partial v \partial r}+2r\frac{\partial\Phi}{\partial v}&+2rf\frac{\partial\Phi}{\partial r}+r^2\frac{\partial f}{\partial r}\frac{\partial\Phi}{\partial r}\\
        &+r^2f\frac{\partial^2\Phi}{\partial r^2}-\mu^2\Phi r^2=0~.
    \end{split}
\end{equation}

Equation (\ref{KG}), giving the dynamics of the massive scalar field $\Phi$, is coupled to equation (\ref{mass}) describing how the mass function $m(v)$ of the black hole  behaves.

In order to study the asymptotic behaviour of the mass function $m(v)$ near the CH, we need to solve the two coupled, inhomogeneous, nonlinear, partial differential equations, given by equations (\ref{mass}) and (\ref{KG}) in the near horizon and long time limit.

\section{Asymptotic Dynamics}
\label{sec-3}
    We can simplify the coupled equations by assuming that the massive scalar field, $\Phi(r,v)$, is spherically symmetric (S wave approximation) so that its intensity falls off like $1/r^2$, and we write 
\begin{equation}
\label{Phi}
    \Phi(r,v)=\frac{1}{r}\phi(r,v)
\end{equation}
Substituting equation (\ref{Phi}) in equations (\ref{mass}) and (\ref{KG}), we obtain
\begin{equation}
\label{modified mass and KG equation}
    \begin{split}
        &\frac{dm}{dv} =4\pi\left[\left( \frac{\partial\phi}{\partial v} \right)^2+\frac{1}{2}\mu^2f\phi^2\right.\\
        &\hspace{15mm}\left.+f\frac{\partial\phi}{\partial v}\left(\frac{\partial\phi}{\partial r}-\frac{\phi}{r}\right)+\frac{1}{2}f^2\left(\frac{\partial\phi}{\partial r}-\frac{\phi}{r}\right)^2\right]\\
        \end{split}
\end{equation}
and 
\begin{equation}
    \label{KG11}
    \begin{split}
         &rf\frac{\partial^2\phi}{\partial r^2}+2r\frac{\partial^2\phi}{\partial v\partial r}+r\frac{\partial f}{\partial r}\frac{\partial \phi}{\partial r}-\phi\frac{\partial f}{\partial r}-\mu^2\phi r=0~.\\
    \end{split}
\end{equation}

To explore the dynamics near the CH, we expand  $f(r,v)$ and $\phi(r,v)$ about the inner horizon, defined by $r_0=m_0-\sqrt{m_0^2-Q^2}$,
\begin{equation}
    \label{series expansion}
    \begin{split}
        f(r,v) &=f_0(v)+x f_1(v)+x^2 f_2
(v)+\dots\\
        \phi(r,v) &=\phi_0(v)+x\phi_1(v)+x^2\phi_2(v)+\dots\\
    \end{split}
\end{equation}
where  $x=r-r_0$.
    In the near horizon approximation $x\xrightarrow[]{}0$, we can neglect the contributions from $f_2$, as it is expected to be much smaller compared to the first-order contribution from $f_1$. The important point to note here is that we are only interested in obtaining the behaviour of the black hole when the infalling ovserver is in between the two horizons, where $f(r,v)$ is negative. Thus
    $$f_0(v)=\frac{2m(v)}{r_0}$$
    and
    $$f_1(v)=-\frac{2}{r_0^2}\left[m_0+m(v)-\frac{Q^2}{r_0}\right]~.$$

Now substituting the series expansions from equations (\ref{series expansion}) into equations (\ref{modified mass and KG equation}) and (\ref{KG11}) yeilds 

\begin{equation}
    \label{eq aftr srs expnsn}
    \begin{split}
    \frac{dm}{dv}& =4\pi\left[ \left(\frac{d\phi_0}{dv} \right)^2+\frac{1}{2}f_0\mu^2\phi_0^2\right.\\
    &+f_0\frac{d\phi_0}{dv}\left(\phi_1-\frac{\phi_0}{r_0} \right)
    \left.+ \frac{1}{2}f_0^2\left(\phi_1-\frac{\phi_0}{r_0} \right)^2      \right]
    \end{split}
\end{equation}
and 
\begin{equation}
    \label{eq after srs expnsn KG}
    \frac{d\phi_1}{dv}+f_0\phi_2+\frac{f_1}{2}\left(\phi_1-\frac{\phi_0}{r_0} \right)-\frac{1}{2}\mu^2\phi_0=0~,
\end{equation}
at zeroth order in $x$. And from equation (\ref{KG11})

\begin{equation}
\label{phi2 eqn}
    \begin{split}
        \frac{d\phi_2}{dv}+f_1\phi_2=\frac{1}{4}\left[\mu^2\phi_1+\frac{f_1}{r_0}\left(    \phi_1-\frac{\phi_0}{r_0} \right)     \right]
    \end{split}
\end{equation}
at first order in $x$, which can be re-expressed as
    \begin{equation*}
    \begin{split}
        4r_0&\frac{d}{dv}\left[\left(\phi_2+\frac{\phi_0}{4r_0^2}\right)\exp\left(\int f_1 dv\right)  \right]\\
        &=e^{-\mu^2r_0v}\frac{d}{dv}
        \left[\phi_1\exp\left(\int(\mu^2r_0+f_1)dv  \right)\right]\\
        &-\exp\left( \int f_1dv  \right)\frac{d}{dv}\left( \phi_1-\frac{\phi_0}{r_0}  \right)~.
    \end{split}
\end{equation*}
In the Cauchy horizon limit, $v\xrightarrow[]{}\infty$, the first term on the right hand side vanishes, and we obtain                          
\begin{equation}
\label{closure}
    \begin{split}
        4r_0&\frac{d}{dv}\left[\left(\phi_2+\frac{\phi_0}{4r_0^2}\right)\exp\left(\int f_1 dv\right)  \right]\\
        &=-\exp\left( \int f_1dv  \right)
        \frac{d}{dv}\left( \phi_1-\frac{\phi_0}{r_0}  \right)~.
    \end{split}
\end{equation}

The nonlinearity of the original equations (\ref{modified mass and KG equation}) and (\ref{KG11}) introduces a closure problem in the above perturbed equations. This closure problem can be circumvented by balancing equation (\ref{closure}) with the approximations $\phi_2\approx-\phi_0/(4r_0^2)$ and $\phi_1\approx\phi_0/r_0$.
These approximations are expected to work well for the near extremal case $m_0\sim Q$. Thus the series expansion for $\phi(r,v)$ takes the form
\begin{equation}
\label{sclr fild srs expnsn soln}
    \phi(r,v)\bigg|_\text{near CH}=\phi_0\left[1+\frac{x}{r_0}-\frac{x^2}{4r_0^2}+\dots           \right]~.
\end{equation}
    
With these approximations, equation (\ref{eq after srs expnsn KG}) gives  
\begin{equation}
    \begin{split}
        & \frac{d\phi_0}{dv}=\frac{1}{2}\Biggl(r_0\mu^2+\frac{m(v)}{r_0^2}\Biggl)\phi_0\\
    \end{split}
\end{equation}
in the CH limit.
Solution of this equation is readily obtained as 
\begin{equation}
    \label{phi_0 solution}
    \phi_0(v)=A\phantom{0}e^{\mu^2r_0v/2}\exp\bigg(\frac{1}{2r_0^2}\int mdv  \bigg)~,
\end{equation}
where $A$ is the integration constant. Substituting this solution in the mass equation (\ref{eq aftr srs expnsn}) leads to 
\begin{equation}
\label{integro}
    \begin{split}
    \frac{r_0^4}{A^2\pi}&\exp\left(-\int \frac{m}{r_0^2}dv   \right)\frac{dm}{dv}\\
    &=e^{\mu^2r_0v}\big( m^2+6\mu^2r_0^3m+r_0^6\mu^4  \big)
    \end{split}
\end{equation}
This last mass equation (\ref{integro}) is in the form of an integro-differential equation. For the sake of convenience, we reexpress it as a second order differential equation of the form

\begin{equation}
\label{finl mass equn}
    \begin{split}
    \frac{d^2h}{dv^2}=e^{h+av}\biggl[ \bigg( \frac{dh}{dv} \bigg)^2 +6a\frac{dh}{dv}+a^2   \biggl]
    \end{split}
\end{equation}
with $a=\mu^2r_0$, $A^2=r_0^4/\pi$, and $$h(v)= \frac{1}{r_0^2}\int m(v)\, dv,$$ 
so that
$$m=r_0^2\frac{dh}{dv}~.$$

To solve the second order differential equation (\ref{finl mass equn}), we employ the standard Adomian decomposition method (ADM) with the initial condition that $h(v_0)=\alpha>0$ and $h'(v_0)=\beta>0$, where $v_0$ is assumed to be the moment of crossing the outer horizon.

\section{Adomian Decomposition Method}
\label{sec-4}
In this section, we give a brief account of the  Adomian decomposition method (ADM). For more detailed discussions, we invite the reader to Refs. \cite{adomian1990review, adomian1988review, adomian2013solving, adomian2014nonlinear, duan2012review}, and references therein.

The ADM is a perturbative method, useful for solving a variety of ordinary and partial linear or nonlinear (homogeneous or inhomogeneous) differential equations. The solution for the dependent variable, $y(x_1,x_2,\dots, x_n)$ is given in terms of an infinite series, which converges if an analytic solution exists.

Consider the general inhomogeneous, nonlinear ordinary differential equation,
\begin{equation}
\label{general form of inhomgnos nonlnr ode}
    L[y]+N[y]+R[y]=g(x)~,
\end{equation}
where $L$ is the highest order linear differential operator, $N$ is the nonlinear part of the equation and $R$ is the remaining linear part, and $g$ is the source. If $L=\frac{d^k}{dx^k},$ assuming that the inverse of $L$  exists, then $L^{-1}$ is defined as
\begin{equation*}
    L^{-1}=\int_{x_0}^x dx \int_{x_0}^x dx\dots \int_{x_0}^x dx \phantom{000}(\text{$k$ times})~,
\end{equation*}
where $x_0$ is the initial value of $x$.

Applying $L^{-1}$ on equation (\ref{general form of inhomgnos nonlnr ode}) yeilds
\begin{equation}
\label{l inverse}
    L^{-1}L[y]=L^{-1}g-L^{-1}N[y]-L^{-1}R[y]~.
\end{equation}

The ADM prescribes that the dependent variable $y$ can be written as an infinite sum of the form
\begin{equation}
\label{y seriess}
    y(x)=\sum_{n=0}^\infty y_n(x)~,
\end{equation}
and the nonlinear part $N[y]$ can be decomposed in terms of a series of polynomials, known as Adomian polynomials, such that
\begin{equation*}
    \begin{split}
        N[y]=\sum_{n=0}^\infty A_n~,
    \end{split}
\end{equation*}
with
\begin{equation}
    A_n=\frac{1}{n!}\frac{d^n}{d\lambda^n}\left[N\left(\sum_{p=0}^\infty y_p\lambda^p  \right)    \right]\Bigg|_{\lambda=0}~.
\end{equation}
Putting these summations in equation (\ref{l inverse}) leads to 
\begin{equation}
\begin{split}
    L^{-1} L  \sum_{n=0}^\infty y_n(x)=L^{-1}g-&L^{-1}\bigg( R\sum_{n=0}^\infty \big[y_n(x)\big] \bigg)\\
    &-L^{-1}\bigg( \sum_{n=0}^\infty A_n \bigg)~.
\end{split}  
\end{equation}

In our case, $L$ is of second order $L=\frac{d^2}{dx^2}$, and we get the recurrence relations
\begin{equation}
    \begin{split}
        \label{recurrence algorithm}
        & y_0=y(x_0)+y'(x_0)(x-x_0)+L^{-1}g\\
        & y_1=-L^{-1}(y_0))-L^{-1} R(y_0)-L^{-1}(A_0)\\
        & \dots,\\
        &y_{n+1}=-L^{-1}(y_n)-L^{-1} R(y_n)-L^{-1}(A_n)~.
    \end{split}
\end{equation}

 In some cases, the nonlinear part may have a very complicated structure and calculation of the Adomian polynomials 
$(A_n)$ become very cumbersome.  In such cases, $A_n$ can be calculated using a Taylor series. Supposing the nonlinear part is some function of $y$ denoted by $F(y)$, the Taylor expansion of $F(y)$ around $y_0$ is given by 
\begin{equation*}
    \begin{split}
    F(y)=F(y_0)&+F'(y_0)(y-y_0)\\
    &+\frac{1}{2!}F''(y_0)(y-y_0)^2+\dots~.
    \end{split}
\end{equation*}
Upon substituting the series expansion (\ref{y seriess}) for $y(x)$ in the above Taylor expansion yields
\begin{equation*}
         \begin{split}
          F(y)=F(y_0)&+F'(y_0)(y_1+y_2+...)\\
          &+\frac{1}{2!}F''(y_0)(y_1+y_2+...)^2+...
          \end{split}
\end{equation*}
The required Adomian polynomials can be obtained from the above Taylor expansion as
\begin{equation}
\label{adomian polynmls}
\begin{split}
    & A_0 =F(y_0)\\
    & A_1 =F'(y_0)y_1\\
    & A_2 =F'(y_0)y_2+\frac{1}{2!}F'(y_0)y_1^2\\
    & A_3 = F'(y_0)y_3+\frac{2}{2!}F''(y_0)y_1y_2+\frac{1}{3!}F'''(y_0)y_1^3
\end{split}
\end{equation}

We shall utilise this methodology in the next section to solve our differential equation (\ref{finl mass equn}).

\section{Solution using ADM }
\label{sec-5}
It is easy to see from equations (\ref{finl mass equn}) and (\ref{general form of inhomgnos nonlnr ode}) that $R=0$ and $g=0$ in our present case. The nonlinear part is given by the function $$F=e^{h+av}[(h')^2+6ah'+a^2]$$ and $$L=\frac{d^2}{dv^2} ~ \text{giving}~ L^{-1}=\int_{v_0}^v dv\int_{v_0}^v dv.$$ 

 As stated earlier, we have the initial conditions $h[v_0]=\alpha$ and $h'[v_0]=\beta=m_0/r_0^2$, and the mass function $m(v)$ is related to $h(v)$ by $$m(v)=r_0^2\frac{dh}{dv}=r_0^2\frac{d}{dv}\left[\sum_{n=0}^\infty h_n(v)\right]. $$ Thus from equation (\ref{finl mass equn}), we have
\begin{equation*}
\begin{split}
        L^{-1}L[h]=L^{-1}N[h]~,
\end{split}
\end{equation*}
giving $$ h[v]=h[v_0]+(v-v_0)h'[v_0]+L^{-1}\left[\sum_{n=0}^\infty A_n\right]~.$$
We therefore obtain $h_0(v)=\alpha+\beta(v-v_0)$, and the remaining $h_n$'s are given by the recurrence relation 
\begin{equation*}
    h_{n+1}(v)=L^{-1}[A_n] \phantom{0} , \hspace{10mm} n\geq0
\end{equation*}
where the Adomian polynomials, $A_n$, can be calculated using equation (\ref{adomian polynmls}). Some of the first few $h_n(v)$ are listed below:

\begin{equation}
    \begin{split}
    \label{solun for h}
         h_0(v)&= \alpha+\beta(v-v_0)\\
         h_1(v)&=e^{(\alpha-\beta v_0)}(a^2+6a\beta+\beta^2)\left[e^{v(a+\beta)}\right.\\
        &\left.+e^{v_0(a+\beta)}\left(-1-(v-v_0)(a+\beta)   \right)\right]\\
         h_2(v)&= \frac{e^{2(\alpha-\beta v_0)}(a^2+6a\beta+\beta^2)^2}{4\beta(a+\beta)^3}\left[ e^{2v(a+\beta)} \right.\\
        &+e^{(v+v_0)(a+\beta)}\{4-4(v-v_0)(a+\beta)    \}\\
        &\left.+e^{2v_0(a+\beta)}\{-5-2(v-v_0)(a+\beta)\}                              \right]\\
        h_3(v) &=\frac{e^{3(\alpha-\beta v_0)}(a^2+6a\beta+\beta^2)^3}{12\beta^2(a+\beta)^4}\Biggl[e^{3v(a+\beta)}\\ 
        &+e^{3v_0(a+\beta)}\{-22-6(v-v_0)(a+\beta)\}\\
        &+e^{(2v+v_0)(a+\beta)}\{6-6(v-v_0)(a+\beta)\}\\
        &+3e^{(v+2v_0)(a+\beta)}\Biggl\{5+2(v-v_0)(a+\beta)\\
        &\times\bigg(-3+(v-v_0)(a+\beta)\bigg) \Biggl\}\Biggl]\\
        \end{split}
\end{equation}

The corresponding mass function is readily obtained as 
\begin{equation*}
    \begin{split}
        \frac{m(v)}{r_0^2}&=\Bigg[ h_o'(v)+h_1'(v)+h_2'(v)+h_3'(v)\dots\Bigg]\\
        & =\beta +\frac{1}{\sigma}\Bigg[\rho\sigma(v-v_0)e^{(\delta+v)}\\
        & \phantom{0000}+\frac{\rho^2\sigma^2}{6}(v-v_0)^3e^{2(\delta+v)}\\
        &\phantom{0000}+\frac{\rho^3\sigma^3}{30}(v-v_0)^5 e^{3(\delta+v)}+\dots \Bigg]~,
    \end{split}
\end{equation*}
which can be written as
\begin{equation}
    \label{mass solution}
    \frac{m}{r_0^2}=\beta+\frac{1}{\sigma}+\frac{1}{\sigma(v-v_0)}\sum_{n=0}^\infty \frac{z^n}{n!(2n-1)}~,
\end{equation}
where $z=\rho\sigma(v-v_0)^2\exp{(\delta+\beta v)},$ with $\gamma=a+\beta, \  \delta=\alpha-\beta v_0,\  \rho=(a^2+6a\beta+\beta^2),$ and $\sigma=\gamma/\beta$.

A ratio test of the series (\ref{mass solution}) shows that it is  convergent for any value of $z$.
The summation can be approximated as
\begin{equation}
\label{summation approximation}
  \sum_{n=0}^\infty \frac{z^n}{n!(2n-1)}=\frac{4}{z^2}(e^z-1)-\frac{4}{z}-3+\frac{z}{3}~,  
\end{equation}
which is an excellent approximation because the series expansion terms on the right hand side $(b_n z^n)$ differ from those on the left hand side $(a_n z^n)$ within $|b_n-a_n|\sim 10^{-n-3}$.

Thus the mass function of the black hole,  as measured by an infalling observer, is obtained from equations (\ref{mass solution}) and equation (\ref{summation approximation}) as

\begin{equation}
    \label{final mass solution}
    \begin{split}
            &\frac{m(v)-m_0}{r_0^2}\\
            &\hspace{8mm}\approx \frac{4\exp{\left[(v-v_0)^2\rho\sigma e^{(\delta+\gamma v)}
            -2(\delta+\gamma v)\right]}}{(v-v_0)^{5}\rho^2\sigma^2}\\
            &\hspace{15mm}+\frac{1}{3}(v-v_0)\rho e^{(\delta+\gamma v)}\\
    \end{split}
\end{equation}
where only the dominant contributions have been included. 

The first term on the right hand side of equation (\ref{final mass solution}) gives the most dominant contribution because of its double exponential nature. Consequently, the CH instability is enormously enhanced in the limit $v\xrightarrow[]{}\infty$.
\begin{figure}[H]
  \centering
    \includegraphics[scale=0.3]{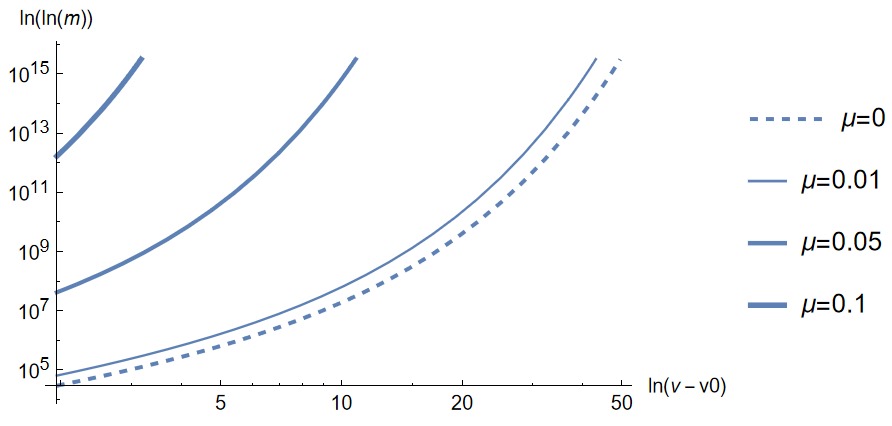}
    \caption{Plot displaying mass superinflation with different values of scalar mass $\mu$.  } 
    \label{mass inflation}
\end{figure}
\section{Discussion and Conclusion}
\label{sec-6}
 In this work, we considered the dynamics of a massive scalar field coupled to the dynamics of the mass function of the Reissner-Nordstr\"om black hole to find the nature of Cauchy horizon (CH) singularity. The scalar field $\Phi(r,v)$ was assumed to be chargeless and spherically symmetric with its intensity falling off like $1/r^2$.
As a result of this perturbation, the redshift function modifies to $f=1-2[m_0+m(v)]/r+Q^2/r^2$. Employing the Einstein-Maxwell field equation coupled with the Klein-Gordon equation, we obtained the coupled dynamics of the mass function $m(v)$ and the scalar field $\Phi(r,v)$ given by equations (\ref{mass}) and (\ref{KG}). To solve the coupled set of nonlinear differential equations, we developed a perturbative method and obtained a second order nonlinear differential equation given by equation (\ref{finl mass equn}). To solve this equation, we employed the Adomian decomposition method (ADM). This results in a double exponential behaviour for the mass function $m(v)$ in the asymptotic limit, as shown in equation (\ref{final mass solution}).      

It is evident from  equation (\ref{final mass solution}) that  the mass of the Reissner-Nordstr\"om black hole, as measured by an infalling observer, shows a rapid and unbounded double exponential growth upon approaching the CH. 

 Figure \ref{mass inflation} displays the behaviour of the mass function for different values of scalar mass $\mu$ including the massless case. The plot illustrates that the mass inflation in the massive case is stronger than in the massless case,  the behaviour being dictated by the double exponential nature in both cases. This is also evident from the expression for the mass function $m(v)$ given by equation (\ref{final mass solution}) [where the coefficients are  $\rho=(a^2+6a\beta+\beta^2)$ and $\gamma=a+\beta$, with $a=\mu^2 r_0^2$, $\sigma=\gamma/\beta$, $\beta=m_0/r_0^2$ and $\delta=\alpha-\beta v_0$ is a constant]. It is important to note that such double exponential growth is a enormously stronger mass inflation compared to those obtained previous studies.

In the near horizon and long time limits ($ x\xrightarrow[]{}0$ and $v\xrightarrow[]{}\infty$), the dominant behaviour of the mass function and the scalar field can be expressed as 
$$m(v)\sim v^{-5}e^{-2\gamma v}\exp{[v^2\rho\sigma e^{\gamma v}]}$$ 
and 
$$\phi(v)\sim \exp\bigg[\frac{1}{2r_0^2}\int m(v)\,dv\bigg].$$
 The strongly diverging nature of the scalar field at the CH comes from the strong gravitational blueshift of the continuous scalar field surrounding the black hole.  Because of this blueshift, the scalar field rapidly accumulates at the CH with increasing energy  which, in turn, results in a rapid \textit{double exponential} growth in the mass of the black hole, creating what may be termed as \textit{mass superinflation}. This implies that the information accumulated at the inner horizon would be seen by the  infalling observer and all the history of the external universe containing the scalar field will be available to him/her within a fraction of time, as was first predicted by Penrose \cite{Schpf1970BattelleR}. Because of the double exponentially diverging mass function, the Kretschmann scalar $R_{\mu\nu\alpha\beta}R^{\mu\nu\alpha\beta}\propto m^2$ is also a double exponentially diverging function, confirming an enormously strong singularity at the CH. Such strong divergence suggests that there is no possibility of extending the spacetime beyond the Cauchy horizon.

\section{Acknowledgement}
Nihar Ranjan Ghosh is supported by a Research  Fellowship of the Ministry of Education (MoE), Government of India.

 \bibliographystyle{unsrt}

\begin{thebibliography}{10}

\bibitem{reissner1916eigengravitation}
Hans Reissner.
\newblock {\"U}ber die eigengravitation des elektrischen feldes nach der
  einsteinschen theorie.
\newblock {\em Annalen der Physik}, 355(9):106--120, 1916.

\bibitem{nordstrom1918energy}
G~Nordstrom.
\newblock On the energy of the gravitational field in einstein’s theory ii
  verhandl.
\newblock {\em Koninkl. Ned. Akad. Wetenschap., Afdel. Natuurk., Amsterdam},
  26:1201--1208, 1918.

\bibitem{penrose1969gravitational}
Roger Penrose.
\newblock Gravitational collapse: The role of general relativity.
\newblock {\em Nuovo Cimento Rivista Serie}, 1:252, 1969.

\bibitem{novikov1966change}
ID~Novikov.
\newblock Change of relativistic collapse into anticollapse and kinematics of a
  charged sphere.
\newblock {\em JETP Lett.(USSR)(Engl. Transl.)}, 3, 1966.

\bibitem{simpson1973internal}
Michael Simpson and Roger Penrose.
\newblock Internal instability in a reissner-nordstr{\"o}m black hole.
\newblock {\em International Journal of Theoretical Physics}, 7:183--197, 1973.

\bibitem{Schpf1970BattelleR}
R.~Penrose.
\newblock Battelle rencontres ‐ 1967 lectures in mathematics and physics (ed.
  by c. m. dewitt and j. a. wheeler). xvii + 557 s. m. fig. new york/amsterdam
  1968. w. a. benjamin, inc.
\newblock {\em Zamm-zeitschrift Fur Angewandte Mathematik Und Mechanik},
  50:782--782, 1970.

\bibitem{ori1991inner}
Amos Ori.
\newblock Inner structure of a charged black hole: An exact mass-inflation
  solution.
\newblock {\em Physical review letters}, 67(7):789, 1991.

\bibitem{hamilton2010physics}
Andrew~JS Hamilton and Pedro~P Avelino.
\newblock The physics of the relativistic counter-streaming instability that
  drives mass inflation inside black holes.
\newblock {\em Physics reports}, 495(1):1--32, 2010.

\bibitem{poisson1989inner}
Eric Poisson and Werner Israel.
\newblock Inner-horizon instability and mass inflation in black holes.
\newblock {\em Physical review letters}, 63(16):1663, 1989.

\bibitem{poisson1990internal}
Werner Israel and Eric Poisson.
\newblock Internal structure of black holes.
\newblock {\em Physical Review D}, 41(6):1796, 1990.

\bibitem{brown2011mass}
Eric~G Brown, Robert Mann, and Leonardo Modesto.
\newblock Mass inflation in the loop black hole.
\newblock {\em Physical Review D}, 84(10):104041, 2011.

\bibitem{price1972nonspherical}
Richard~H Price.
\newblock Nonspherical perturbations of relativistic gravitational collapse. i.
  scalar and gravitational perturbations.
\newblock {\em Physical Review D}, 5(10):2419, 1972.

\bibitem{tipler1977singularities}
Frank~J Tipler.
\newblock Singularities in conformally flat spacetimes.
\newblock {\em Physics Letters A}, 64(1):8--10, 1977.

\bibitem{ellis1977singular}
George~FR Ellis and Bernd~G Schmidt.
\newblock Singular space-times.
\newblock {\em General Relativity and Gravitation}, 8:915--953, 1977.

\bibitem{burko1998analytic}
Lior~M Burko and Amos Ori.
\newblock Analytic study of the null singularity inside spherical charged black
  holes.
\newblock {\em Physical Review D}, 57(12):R7084, 1998.

\bibitem{dafermos2017interior}
M~Dafermos and J~Luk.
\newblock The interior of dynamical vacuum black holes i: the, 2017.

\bibitem{davies1976energy}
Paul~CW Davies, Stephen~A Fulling, and William~G Unruh.
\newblock Energy-momentum tensor near an evaporating black hole.
\newblock {\em Physical Review D}, 13(10):2720, 1976.

\bibitem{hiscock1977stress}
William~A Hiscock.
\newblock Stress-energy tensor near a charged, rotating, evaporating black
  hole.
\newblock {\em Physical Review D}, 15(10):3054, 1977.

\bibitem{balbinot1993mass}
Roberto Balbinot and Eric Poisson.
\newblock Mass inflation: The semiclassical regime.
\newblock {\em Physical review letters}, 70(1):13, 1993.

\bibitem{birrell1984quantum}
Nicholas~David Birrell and Paul Charles~William Davies.
\newblock {\em Quantum fields in curved space}.
\newblock Cambridge university press, 1984.

\bibitem{hwang2011internal}
Dong-il Hwang and Dong-han Yeom.
\newblock Internal structure of charged black holes.
\newblock {\em Physical Review D}, 84(6):064020, 2011.

\bibitem{barcelo2021black}
Carlos Barcel{\'o}, Valentin Boyanov, Ra{\'u}l Carballo-Rubio, and Luis~J
  Garay.
\newblock Black hole inner horizon evaporation in semiclassical gravity.
\newblock {\em Classical and Quantum Gravity}, 38(12):125003, 2021.

\bibitem{barcelo2022classical}
Carlos Barcel{\'o}, Valentin Boyanov, Ra{\'u}l Carballo-Rubio, and L~J Garay.
\newblock Classical mass inflation versus semiclassical inner horizon
  inflation.
\newblock {\em Physical Review D}, 106(12):124006, 2022.

\bibitem{carballo2022regular}
Ra{\'u}l Carballo-Rubio, Francesco Di~Filippo, Stefano Liberati, Costantino
  Pacilio, and Matt Visser.
\newblock Regular black holes without mass inflation instability.
\newblock {\em Journal of High Energy Physics}, 2022(9):1--14, 2022.

\bibitem{franzin2022stable}
Edgardo Franzin, Stefano Liberati, Jacopo Mazza, and Vania Vellucci.
\newblock Stable rotating regular black holes.
\newblock {\em Physical Review D}, 106(10):104060, 2022.

\bibitem{mcmaken2023semiclassical}
Tyler McMaken.
\newblock Semiclassical instability of inner-extremal regular black holes.
\newblock {\em Physical Review D}, 107(12):125023, 2023.

\bibitem{brady1992cauchy}
PR~Brady and E~Poisson.
\newblock Cauchy horizon instability for reissner-nordstrom black holes in de
  sitter space.
\newblock {\em Classical and Quantum Gravity}, 9(1):121, 1992.

\bibitem{mellor1990stability}
Felicity Mellor and Ian Moss.
\newblock Stability of black holes in de sitter space.
\newblock {\em Physical Review D}, 41(2):403, 1990.

\bibitem{brady1993cauchy}
Patrick~R Brady, Dario Nunez, and Sukanya Sinha.
\newblock Cauchy horizon singularity without mass inflation.
\newblock {\em Physical Review D}, 47(10):4239, 1993.

\bibitem{cai1995black}
Rong-Gen Cai and Ru-Keng Su.
\newblock Black holes in de sitter space and the stability conjecture of cauchy
  horizons.
\newblock {\em Physical Review D}, 52(2):666, 1995.

\bibitem{helliwell1993testing}
TM~Helliwell and DA~Konkowski.
\newblock Testing a stability conjecture for cauchy horizons.
\newblock {\em Physical Review D}, 47(10):4322, 1993.

\bibitem{konkowski1994instabilities}
DA~Konkowski and TM~Helliwell.
\newblock Instabilities of the cauchy horizon in kerr black holes.
\newblock {\em Physical Review D}, 50(2):841, 1994.

\bibitem{markovic1995classical}
Dragoljub Markovi{\'c} and Eric Poisson.
\newblock Classical stability and quantum instability of black-hole cauchy
  horizons.
\newblock {\em Physical review letters}, 74(8):1280, 1995.

\bibitem{adomian1990review}
G~Adomian.
\newblock A review of the decomposition method and some recent results for
  nonlinear equations.
\newblock {\em Mathematical and Computer Modelling}, 13(7):17--43, 1990.

\bibitem{adomian1988review}
George Adomian.
\newblock A review of the decomposition method in applied mathematics.
\newblock {\em Journal of mathematical analysis and applications},
  135(2):501--544, 1988.

\bibitem{adomian2013solving}
George Adomian.
\newblock {\em Solving frontier problems of physics: the decomposition method},
  volume~60.
\newblock Springer Science \& Business Media, 2013.

\bibitem{adomian2014nonlinear}
George Adomian.
\newblock {\em Nonlinear stochastic operator equations}.
\newblock Academic press, 2014.

\bibitem{duan2012review}
Jun-Sheng Duan, Randolph Rach, Dumitru Baleanu, and Abdul-Majid Wazwaz.
\newblock A review of the adomian decomposition method and its applications to
  fractional differential equations.
\newblock {\em Communications in Fractional Calculus}, 3(2):73--99, 2012.

\end{thebibliography}

\end{multicols}
\end{document}